\documentclass[12pt,leqno]{article}

\usepackage{graphicx}

 \font\gotb eufm10 scaled \magstep1
\newcommand{\bb}{\bibitem}
\newcommand{\cc}{\cite}
\newcommand{\vp}{\varphi}
\newcommand{\sss}{\sigma}
\newcommand{\al}{\alpha}
\newcommand{\Om}{\Omega}

\newcommand{\lt}{\left}
\newcommand{\rt}{\right}
\newcommand{\lll}{\lambda}
\newcommand{\F}{{\cal F}}

\newcommand{\D}{\hat D}

\newcommand{\I}{\hat I}
\newcommand{\0}{\hat 0}

\newcommand{\A}{\hat A}
\newcommand{\B}{\hat B}
\newcommand{\U}{\hat U}
\newcommand{\V}{\hat V}
\newcommand{\W}{\hat W}
\newcommand{\p}{\hat p}

\newcommand{\AAA}{\hbox{\gotb A}}

\newcommand{\HHH}{\hbox{\gotb H}}
\newcommand{\QQ}{\hbox{\gotb Q}}
\newcommand {\QQQ}{\hbox {\gotb Q}_{\xi}}
\newcommand {\qqq}{\hbox {\gotb Q}_{\xi'}}
\newcommand {\vx}{\vp_{\xi}}
\newcommand {\vpx}{\vx(\A)}

\newcommand{\bea}{\begin{eqnarray} \label}
\newcommand{\eeq}{\end{equation}}
\newcommand{\beq}{\begin{equation} \label}
\newcommand{\eea}{\end{eqnarray}}

\newcommand{\rr}[1]{(\ref{#1})}

 \author{D.A Slavnov}

\title{THE POSSIBILITY OF RECONCILING QUANTUM MECHANICS
WITH CLASSICAL PROBABILITY THEORY\thanks{The paper is published in
Theoretical and Mathematical Physics, 149(3): 1689 (2006)} }

   \date{Address: Department of Physics, Moscow State University,
Moscow 119992, Russia. E-mail: slavnov@goa.bog.msu.ru}

\begin{document}

  \maketitle

 \begin{abstract}

We describe a scheme for constructing quantum mechanics in which a
quantum system is considered as a collection of open classical
subsystems. This allows using the formal classical logic and
classical probability theory in quantum mechanics. Our approach
nevertheless allows completely reproducing the standard
mathematical formalism of quantum mechanics and identifying its
applicability limits. We especially attend to the quantum state
reduction problem.

\end{abstract}


\section{Introduction}

In creating the new theory, the pioneers of quantum mechanics were
based on very few experimental facts. As a result, quantum
mechanics proved more a mathematical scheme (albeit quite
successful) than a physical model. It is no accident that the
version of quantum theory proposed by Heisenberg was called
"matrix mechanics": the key notion of a physical theory was
reduced to a purely mathematical object, a matrix.

Schrodinger's wave mechanics was designed as a physical model.
Schrodinger himself persistently attempted to give the main notion
in this model, the wave function, a physical meaning. But the
result was nil. Eventually, it had to be accepted that the wave
function is a probability amplitude. At best, it provides a useful
mathematical object, devoid of direct physical interpretation.

This trend in the development of quantum mechanics reached its
logical height in the famous book by von Neumann~\cc{von}. The
entire role of quantum mechanics was actually reduced to a
physical interpretation of the Hilbert space theory.

On one hand, this allowed providing quantum mechanics with a clear
mathematical structure and developing a powerful mathematical
formalism. Using this formalism allowed describing a huge number
of physical effects mathematically. But on the other hand,
physicists practically abandoned attempts to perceive the physical
nature of quantum phenomena.

Quantum mechanics has become a certain "black box." The initial
data are input into this box. It processes them in accordance with
physically obscure laws and produces an answer, which can then be
compared with experimental results. An excellent agreement follows
in the overwhelming majority of cases.

This normally suffices from the applied standpoint. But on a
larger scale, this is not a problem in natural sciences but rather
a user-manual problem. Not without a reason have the greatest
experts in quantum physics uttered that "one can study quantum
mechanics, can learn how to extract useful results from it, but
cannot understand it."

The dominant role of mathematics in constructing quantum mechanics
is also fraught with danger for the following reason. The desire
to construct a mathematically perfect scheme forces the researcher
to make assumptions and impose conditions that look very natural
from the mathematical standpoint and are simultaneously {\it
sufficient} for correctly describing a physical effect under
consideration (in which case such assumptions are usually
considered to be physical). But whether these mathematical
conditions are {\it necessary} for the physics usually remains
outside the researcher's scope.

Mathematical assumptions can also lead to unexpected physical
corollaries, which have no direct experimental confirmation. For
example, the existence of the trajectory of a quantum particle
contradicts the mathematical formalism of the standard quantum
mechanics. But on the other hand, whenever it proves possible to
trace the motion of a quantum particle in space, this trajectory
is revealed. Of course, more or less reasonable argument is then
offered to explain why the absence of a trajectory cannot be
detected. But there is no direct experimental evidence of this
fact.

The above ideas should in no case be taken as a call for imposing
restrictions on the use of mathematics in physics and in quantum
mechanics in particular. But we must make all possible effort to
ensure that the mathematics follows the physics, rather than vice
versa, in constructing a physical theory. In practical terms, this
means that we must proceed from the physical phenomenon. At the
second stage, when a mathematical description of this phenomenon
is given, we must not only ensure that the mathematical
assumptions are sufficient for such a description but also try to
limit ourselves to only those assumptions that are necessary from
the physical standpoint. Only if the condition of necessity is
satisfied can we be reasonably certain that physical corollaries
of these mathematical assumptions are realized in nature.

Obviously, although the suggested scheme is ideal and is never
realized in its pure form, it must nevertheless be set as a goal.
In formulating quantum mechanics in what follows, we follow this
program as closely as we can.

\section{Observables, measurements, and states}

The fundamental notion of both classical and quantum physics is an
observable. In a physical system, an observable is an attribute
whose numerical value can be obtained using some measuring
procedure. In what follows, we assume that all observables are
dimensionless, which implies fixing some system of units.

In each measurement, the investigated physical system is subject
to the action from the measuring device. Therefore, all
measurements can be divided into two types: reproducible and
nonreproducible. A characteristic feature of reproducible
measurements is then that repeated measurement of the same
observable gives the originally obtained value.

A particularly acute form is taken by the reproducibility question
when several observables are measured in a single physical system.
We assume that we measure an observable  $\A$, then an observable
$\B$, then the observable $\A$ and, again,  finally the observable
$\B$. If we obtain repeated values for each observable as the
result of such repeated measurements, we say that the measurements
of the observables  $\A$ and $\B$ are compatible.

Experiment shows that a key difference between classical and
quantum physical systems is as follows. For classical systems, the
experiment can always be designed such that the measurements of
any two observables are compatible. But a quantum system always
has observables for which a compatible measurement cannot be
realized in any case. Such observables are said to be
incompatible. Accordingly, we say that compatible (or
simultaneously measurable) observables are those for which a
compatible measurement can be made. Devices that allow making
compatible measurements are said to be compatible. For
incompatible observables, no such devices exist.

We let $\AAA_+$ denote the set of all observables in a physical
system under consideration and let $\QQQ$. denote its maximal
subset of compatible observables. It is clear that for a classical
system, this subset coincides with the set $\AAA_+$ itself. For a
quantum system, at least two such subsets must exist. In fact, it
can be verified that they are infinitely many~\cc{slav1}. The
index $\xi$, ranging a set $\Xi$, distinguishes one such subset
from another. A given observable can belong to different subsets
$\QQQ$ simultaneously.

It is easy to verify that each subset  $\QQQ$ can be endowed with
the structure of a real commutative associative algebra. Indeed,
experiment shows that for any compatible observables $\A$ and
$\B$, there exists a third observable $\D$ such that, first, it is
compatible with the observables $\A$ and $\B$, and, second, in
each simultaneous (compatible) measurement of these three
observables, the measurement results are related by

 $$AB=D.$$

Because this relation is satisfied independently of the values of
the individual constituents, it can be assumed that the
observables themselves by definition satisfy the same relation:

   $$\A\B=\D.$$

We can similarly define the addition operation for two compatible
observables and the operation of multiplication of an observable
by a real number.

For a given physical system, using compatible measurements, we
assign each observable  $\A\in\QQQ$ a measurement result:

  $$\A\to A=\vx(\A).$$

This defines a functional $\vx(\A)$ on the algebra  $\QQQ$. By the
definition of the algebraic operations in $\QQQ$, this functional
is a homomorphic map of $\QQQ$ into the set of real numbers. Such
a functional is called a character of a real commutative
associative algebra (see, e.g.,~\cc{dix}).

The characters have the properties

 \beq{4.1}
 \vx(\I)=1,
\eeq
 \beq{4.2}
 \vx(\0)=0,
\eeq
 \beq{4.3}
 \vx(\A^2)\geq 0,
\eeq
 where $\I$, $\0$, and $\A$ are the unit, the zero, and an
arbitrary element of  $\QQQ$.

In each real measurement, we always obtain a finite result.
Therefore, only the observables $\A$ for which the inequality

  $$ \sup_{\xi}\,\sup_{\vx}|\vx(\A)|<\infty.$$
holds can be considered physical.    When this condition is
satisfied, the characters have the additional properties (see,
e.g.,~\cc{rud}):

 \beq{6}
  \vx(\A)=\lll\in \sss(\A;\QQQ),
  \eeq
  \beq{7}
\mbox{if }\lll\in\sss(\A;\QQQ), \mbox{ then there exists a
character } \vx(\A), \mbox{ such that  }\lll= \vx(\A).\eeq

Here,  $\sss(\A,\QQQ)$ is the spectrum of an element  $\A$ in the
algebra $\QQQ$. We recall that the spectrum   $\sss(\A,\AAA)$ of
$\A$ in an algebra  $\AAA$ is the set of numbers $\lll$ such that
the inverse to  $\A-\lll\I$  does not exist in $\AAA$. For each
algebra $\QQQ$, properties~\rr{6} and \rr{7} allow constructing
all of its characters~\cc{slav1}.

The mathematical representation of a physical system is the set of
observables of this system. In what follows, we identify the
physical system with the set of its observables in which relations
between observables are fixed. We identify a subset of observables
with the corresponding physical subsystem. In doing this, we do
not assume that the subsystem must necessarily be somehow isolated
from the rest of the system. The subsystem may not be isolated
spatially and can interact with other parts.

We next discuss the notion of a state of a physical system. We
first consider a classical system. In this case, a state of the
physical system is understood as its attribute that uniquely
predetermines the results of measurements of all the observables.
Mathematically, a state is usually given by a point in phase
space. But it can be easily reckoned that this is just one
specific version of fixing a certain functional on the algebra of
observables, which is a character of this algebra. To be free from
the choice of any specific version, we define a state of a
classical system as a character of the algebra of observables of
this system.

This definition extends to a quantum system as follows. We
consider the set  $\AAA_+$ of observables of a quantum system as a
collection of subsets  $\QQQ$ $(\xi\in\Xi)$, each of which is a
maximal subset of compatible observables. Each of these subsets
has the structure of a real commutative associative algebra and
can be considered the algebra of observables of some classical
subsystem of the quantum system. These classical subsystems are
open, but we can still describe the state of each of them using a
character $\vx(\cdot)$ of the corresponding algebra~$\QQQ$.

We say that an elementary state of a physical system is a
collection   $\vp=[\vx]$ $(\xi\in\Xi)$ of functionals $\vx(\cdot)$
each of which is a character of the corresponding algebra  $\QQQ$.

In each individual measurement, we can measure the observables
belonging to any given $\QQQ$ algebra. The results of such a
measurement are uniquely defined by the corresponding functional
$\vx(\cdot)$ belonging to the collection $\vp$. Thus, the result
of each individual measurement of observables of a physical system
is determined by the elementary state of this system. This
statement holds for both classical systems (in which case the
collection $\vp=[\vx]$ consists of a single functional) and
quantum systems (in which case the collection  $\vp=[\vx]$
consists of infinitely many functionals).

We note that we do not assume the validity of the equality

 \beq{8}
  \vx(\A)=\vp_{\xi'}(\A), \mbox{ if }\A \in \QQQ\cap\qqq.
\eeq Although deceptively natural, this assumption has no
experimental justification, as was shown in~\cc{slav1}.

That condition~\rr{8} is not satisfied implies that a measurement
result can depend not only on the system investigated (on its
elementary state) but also on the type of device used for
measurement. We say that measuring devices belong to the $\xi$
type if for a system in the elementary state $\vp=[\vx]$, the
measurement result for each observable $\A\in\QQQ$ is described by
the functional  $\vx(\cdot)$. This is considered in more detail
in~\cc{slav1}.

We note that many proofs (see, e.g.,~\cc{ksp,ghz}) demonstrating
the impossibility of the existence of physical reality determining
the measurement result are based on tacitly assuming that
conditions of type~\rr{8}.

Obviously, condition~\rr{8}  may be satisfied for some $\vp$.
If~\rr{8} holds for all the  $\QQQ$ containing $\A$, then we say
that the elementary state  $\vp=[\vx]$ is stable for the
observable $\A$.

An elementary state of a quantum system cannot be uniquely fixed
experimentally because the most that can be measured in a single
experiment (or in a group of compatible experiments) is the
observables belonging to one algebra $\QQQ$. In other words, only
values of the functional  $\vx(\cdot)$ can be determined. The
elementary state $\vp$ remains otherwise undefined. For
determining the values of other observables, an additional
experiment must be performed involving a device incompatible with
the one used previously. The new device uncontrollably perturbs
the elementary state that had occurred after the first
measurement. Therefore, the information obtained in the first
experiment becomes obsolete.

In view of this, it is convenient to unite the elementary states
$\vp$ having the same restriction to the algebra $\QQQ$ (i.e., the
functional  $\vx(\cdot)$) into a class  $\vx$-equivalent
elementary states. Thus, only the equivalence class to which the
elementary state of the considered system belongs can be
established in a quantum measurement.

A reproducible measurement of observables belonging to an algebra
$\QQQ$ is reminiscent of the procedure for preparing a quantum
state in the standard quantum mechanics. Accordingly, the class
$\{\vp\}_{\vp_{\xi}}$ of $\vx$-equivalent elementary states $\vp$
that are stable on the subalgebra  $\QQQ$ is said to be a quantum
state $\Psi_{\vp_{\xi}}$.

\section{Probability theory in quantum mechanics}

Defying the widely shared opinion that the Kolmogorov probability
theory~\cc{kol} is inapplicable to quantum systems, we try to use
it (see~\cc{slav1,slav2}). The fundamental notion of the
Kolmogorov probability theory is the probability
space~\cc{kol,nev}) This is a triple  $(\Om,\F, P)$. The first
term in the triple, $\Om$, is a set (space) of elementary events.
The defining properties of elementary events are as follows: (a)
one and only one elementary event occurs in each trial; (b)
elementary events exclude each other. Because two nonorthogonal
quantum states do not exclude each other, they cannot be
elementary events. In our case, the role of an elementary event is
played by the elementary state  $\vp$.

In addition to the elementary event, the notion of an event is
also introduced. Each event  $F$ is identified with some subset of
the set  $\Om$. An event  $F$ is considered to have occurred if
one of the elementary events belonging to this subset ($\vp\in F$)
occurred. The collections of subsets $F$ of the set $\Om$ are
endowed with the structure of a Boolean  $\sss$-algebra.

We recall that the Boolean algebra of a set  $\Om$ is the system
of subsets of this set with three algebraic operations defined on
it: taking the union of subsets, the intersection of subsets, and
the complement of each subset in $\Om$. An algebra is said to be
closed under some algebraic operation if the result of this
operation is an element of this algebra. A Boolean algebra is
called a  $\sss$-algebra if it has the following properties:
first, it contains the set  $\Om$ itself and the empty set
$\emptyset$; second, it is closed under the operation of taking
the complement and a denumerable number of the union and
intersection operations. Accordingly, the second term in the
triple is some Boolean $\sss$-algebra $\F$.

Finally, the third term in the triple is a probability measure
$P$. This is a map of  $\F$ into the set of real numbers (each
$F\in\F$ is sent into a number  $P(F)$) satisfying the conditions
$0\leq P(F) \leq 1$ for all  $F\in\F$, $P(\Om)=1$, and $P(\sum_j
F_j)=\sum_j P(F_j)$ for any denumerable collection of
nonintersecting subsets  $F_j\in \F$. The probability measure is
defined only for the events from the algebra  $\F$. For elementary
events, the probability may not exist in general.

From the physical standpoint, the choice of a  $\sss$-algebra $\F$
is determined by the characteristics of the measuring devices
used. The point is that in reality, measuring devices have a
finite resolving power and therefore cannot always differentiate
one elementary event from another. They can then only be used to
establish that a given experiment involves one of the elementary
events belonging to some subset.

Here is the key difference between classical and quantum physical
systems. In the classical case, we can infinitely increase the
resolving power and use devices that allow simultaneously
measuring the values of an arbitrary number of observables. In the
quantum case, compatible measurements can be performed only for
observables belonging to a given algebra  $\QQQ$. Such
measurements correspond to a certain type of the  $\sss$-algebra,
denoted by $\F_{\xi}$ in what follows. The elements of this
$\sss$-algebra differ in the values (intervals of values) of
observables in the algebra  $\QQQ$. More detailed measurements in
which the values of observables not belonging to $\QQQ$ are
additionally measured are not allowed, because they are
incompatible with the previous measurements. Therefore, the
$\sss$-algebras whose elements additionally differ in the values
of observables not belonging to $\QQQ$ are useless. No probability
measure corresponds to such $\sss$-algebras.

In this connection, we note that it was tacitly assumed in the
proof (see, e.g.,~\cc{chsh})of the famous Bell
inequalities~\cc{bell} that the probability measure always exists.
But just for the quantum systems for which this inequality is
proved, the used probability measures do not exist~\cc{slav2}.
Therefore, the Bell inequalities do not necessarily hold for such
physical systems.

The choice of some  $\sss$-algebra $\F_{\xi}$, mathematically
speaking, makes the set  $\Om$ of elementary events into a
measurable space  $(\Om,\F_{\xi})$. In an experiment, this space
corresponds to a pair: the physical object under investigation and
a certain type of measuring device allowing compatible
measurements of observables from the algebra $\QQQ$.

As a real random variable, we consider a measurable map of a
measurable space  $(\Om,\F_{\xi})$ of elementary events into the
set of real numbers. For an observable $\A$, this takes the form

 $$
\vp\stackrel{\A}{\longrightarrow}A_{\xi}(\vp)\equiv\vx(\A)
\in{\cal R}\equiv[-\infty,+\infty].$$
We note that in the quantum
case, the value of a real random variable can depend not only on
the elementary event (the elementary state) but also on the type
of the measuring device (the index  $\xi$).

We say that a quantum ensemble is a set of physical systems that
are described by the same set $\AAA_+$ of observables and are in
some quantum state. A mixture of quantum ensembles involving each
of these ensembles with some multiplicities is said to be a mixed
quantum ensemble.

Experiment shows that a quantum (mixed) ensemble has probabilistic
properties. It must therefore admit the introduction of a
probability space structure. As a result of a reproducible
measurement, the quantum ensemble passes into a new quantum
ensemble with another probability distribution.

We consider the quantum ensemble of systems that are in a quantum
state  $\Psi_{\vp_{\eta}}$ $(\eta \in \Xi)$. The space
$\Om(\vp_{\eta})$ of elementary events for this ensemble is given
by the equivalence class  $\{\vp\}_{\vp_{\eta}}$. Let a type-$\xi$
device be used in the experiment. This corresponds to a measurable
space $(\Om(\vp_{\eta}),\F_{\xi})$ and a probability measure
$P_{\xi}$.

We measure an observable  $\A\in\QQQ$ and say the event  $F_A$
occurs in the experiment if the registered value of  $\A$ is not
greater than  $A$. Let $P_{\xi}(A)=P(\vp: \vp_{\xi}(\A)\le A)$
denote the probability of this event. If the observable  $\A$ also
belongs to an algebra  $\qqq$, then a $\xi'$-type device could be
used to determining the probability of  $F_A$. In this case, a
different value  $P_{\xi'}(A)$ could be obtained for the
probability. But experiment shows that the same probability is
obtained in this case, i.e.,
 \beq{10}
  P(\vp:\vpx)\le A)= P(\vp:\vp_{\xi'}(\A)\le A).
  \eeq
We introduce the notation

 $$P_{\A}(d\vp)=P(\vp:\vp(\A)\leq A+dA)-P(\vp:\vp(\A)\leq A),$$
where the subscript $\xi$ on the functional  $\vp(\A)$ is omitted
in view of~\rr{10}.

To find the mean of an observable  $\A$ in a quantum state
$\Psi_{\vp_{\eta}}$, we need not consider observables that are
incompatible with $\A$. Therefore, instead of considering the
quantum system, we can restrict ourself to considering its
classical subsystem whose observables are described by the algebra
$\QQQ$ ($\A\in\QQQ$). To determine the mean $\langle\A\rangle$, we
can then use the mathematical formalism of classical probability
theory (see, e.g.,~\cc{nev}) and write

 \beq{12}
\langle\A\rangle=\int_{\vp\in\Psi_{\vp_{\eta}}}\,P_{\A}(d\vp)\,A(\vp)
\equiv\int_{\vp\in\Psi_{\vp_{\eta}}}\,P_{\A}(d\vp)\,\vp(\A)\equiv
\Psi_{\vp_{\eta}}(\A).
  \eeq
Here again, with~\rr{10} in mind, we can drop the subscript $\xi$
on the functionals $\vp(\A)$ and $A(\vp)$ .

Formula~\rr{12} defines a functional  $\Psi_{\vp_{\eta}}(\A)$ (a
quantum mean) on the set  $\AAA_+$. We note that

 \beq{12a}
 \Psi_{\vp_{\eta}}(\I)=\int_{\vp\in\Psi_{\vp_{\eta}}}\,P_{\I}(d\vp)=1.
 \eeq

Obviously,  $\Psi_{\vp_{\eta}}(\al\A)=\al\Psi_{\vp_{\eta}}(\A)$,
where  $\al$ is any real number.   Experiment also shows that for
any  $\A\in\AAA_+$ and $\B\in\AAA_+$, there exists an observable
$\D\in\AAA_+$ such that the relation

$$\Psi{\vp_{\eta}}(\A)+\Psi{\vp_{\eta}}(\B)=\Psi{\vp_{\eta}}(\D).$$
holds for each quantum state $\Psi{\vp_{\eta}}(\cdot)$. Such an
element $\D$  can by definition be considered the sum of  $\A$ and
$\B$.  This means that the set  $\AAA_+$ can be endowed with the
structure of a real linear space such that the
$\Psi{\vp_{\eta}}(\cdot)$ are linear functionals on this space.
Because of property~\rr{4.3}, these functionals are positive.

Because any observable  $\A\in\AAA_+$ is compatible with itself,
it follows that the operation of taking the square of  $\A$ can be
defined on the set $\AAA_+$ following the same scheme as  $\QQQ$.
This allows endowing the linear space  $\AAA_+$ with the structure
of a real Jordan algebra~\cc{jord,emch} with the product of
elements $\A$ and $\B$ defined as

 \beq{14}
 \A\circ\B=1/2\lt((\A+\B)^2-\A^2-\B^2\rt).
  \eeq
This product is manifestly commutative but not associative in
general.

All the Jordan algebras are divided into two classes: special and
exceptional. A Jordan algebra is said to be special if two
conditions are satisfied. First, there exists an associative (not
necessarily real and commutative) algebra  \AAA{} such that the
set  $\AAA_+$ as a linear space is a subspace in \AAA. In the
algebra \AAA{} in addition to the original associative product
$\U\V$, a product can be introduced using formula~\rr{14}, which
in this case becomes

 $$
 \U\circ\V=1/2(\U\V+\V\U).
  $$
With respect to this product, the set  \AAA{} is a Jordan algebra.
Second, as a Jordan algebra, $\AAA_+$ must be a subalgebra
in~$\AAA$.

Whether exceptional Jordan algebras can be used in quantum physics
is unknown. In all quantum models considered to date, the set of
observables can be endowed with the structure of a special Jordan
algebra.

In line with this historical experience, we assume the following
hypothesis.

  \

{\bf Hypothesis.  } {\it There exists an involutive, associative,
and in general noncommutative algebra  \AAA{} satisfying the
following conditions:\\
 a.  For each element  $\U\in\AAA$, there
exists a Hermitian element  $\A$ such that  $\U^*\U=\A^2$.\\
 b.  If $\U^*\U=0$, then   $\U=0$.\\
c. The set of Hermitian elements of the algebra  \AAA{} coincides
with the set  $\AAA_+$ of observables.}

  \

In assuming this hypothesis, we must depart from rigorously
observing the rule to make only those assumptions whose necessity
follows from physical experiment. We must be satisfied with a less
reliable criterion, the historical experience. At the same time,
we emphasize that in the standard quantum mechanics, in addition
to this hypothesis, a much less obvious conjecture is also
adopted: it is assumed that the observables are self-adjoint
operators in some Hilbert space.

In what follows, we can consider a physical system specified if
the algebra  \AAA{} is given. Because the algebras $\QQQ$ of
compatible observables are maximal real commutative subalgebras in
\AAA{} belonging to  $\AAA_+$, it follows that compatible
observables are pairwise commuting elements of  \AAA{}, while
incompatible observables do not commute with each other.

\section {Dynamical variables and a  $C^*$-algebra}

The elements of the algebra  \AAA{} are called dynamical variables
in what follows. Any element  $\U\in \AAA$ is uniquely represented
as  $\U=\A+i\B$, where $\A,\B\in\AAA_+$. Therefore, the functional
$\Psi_{\vp_{\eta}}(\,\cdot\,)$ can be uniquely extended to a
linear functional on  \AAA{} as
$\Psi_{\vp_{\eta}}(\U)=\Psi_{\vp_{\eta}}(\A)+i\Psi_{\vp_{\eta}}(\B)$.

As shown in~\cc{slav1} the equality

 $$
\sup_{\eta}\sup_{\vp_{\eta}}\Psi_{\vp_{\eta}}(\U^*\U)=
\sup_{\eta}\sup_{\vp_{\eta}}\vp_{\eta} (\U^*\U) $$ holds;
therefore (see \cc{slav1,emch}), a norm can be introduced in the
algebra  $\U$ using the equality

 $$
 \|\U\|^2=\sup_{\eta}\,\sup_{\vp_{\eta}}\vp_{\eta}(\U^*\U).
 $$
Because the functional  $\vp_{\eta}$ is a character of
$\QQ_{\eta}$, we have
$\vp_{\eta}([\U^*\U]^2)=[\vp_{\eta}(\U^*\U)]^2$. This implies that

 \beq{17}
\|\U^*\U\|=\|\U\|^2.
 \eeq
 A complete normalized involutive algebra
whose norm satisfies additional condition~\rr{17} is called a
$C^*$-algebra~\cc{dix}. Therefore, the algebra of quantum
dynamical variables can be endowed with the structure of a
$C^*$-algebra.

A remarkable property of  $C^*$-algebras is that any $C^*$-algebra
is isometrically isomorphic to a subalgebra of linear bounded
operators in an appropriate Hilbert space $\HHH$~\cc{dix}. A
faithful representation of  the $C^*$-algebra is said to be
realized in the space  $\HHH$. This allows incorporating the
mathematical formalism of the standard quantum mechanics into the
scheme proposed in this paper.

The relation of a  $C^*$-algebra to a Hilbert space is realized by
the so-called canonical Gelfand-Naimark-Segal (GNS) construction
(see, e.g.,~\cc{naj,emch}). It consists in the following.

Let there be some  $C^*$-algebra \AAA{} and a linear positive
functional  $\Psi$ on it. We consider two elements
$\U,\,\U'\in\AAA$ equivalent if $\Psi\left(\W^*(\U-\U')\right)=0$
for any  $\W\in\AAA$. We let  $\Phi(\U)$ denote the equivalence
class of $\U$ and consider the set  $\AAA(\Psi)$ of all
equivalence classes in  \AAA. We make this set  $\AAA(\Psi)$ into
a linear space by setting $a\Phi(\U)+b\Phi(\V)=\Phi(a\U+b\V)$. In
$\AAA(\Psi)$), we define a scalar product as

\beq{18}
 \left(\Phi(\U),\Phi(\V)\right)=\Psi(\U^*\V).
  \eeq
This scalar product induces the norm $\|\Phi(\U)\| =
[\Psi(\U^*\U)]^{1/2}$ in the algebra  $\AAA(\Psi)$. The completion
with respect to this norm makes $\AAA(\Psi)$ into a Hilbert space.
Each element  $\V$ of \AAA{} is uniquely represented in this space
by a linear operator  $\Pi(\V)$ acting as

  \beq{19}
\Pi(\V)\Phi(\U)=\Phi(\V\U).
  \eeq
The GNS construction thus yields a representation of the
$C^*$-algebra by linear operators in a Hilbert space.

We consider the GNS construction with the functional generating
the representation given by  $\Psi_{\vp_{\eta}}(\B)$. If
$\Phi(\I)$ is the equivalence class of the element  $\I$, then
from~\rr{18} and \rr{19}, we obtain

 \beq{20}
 \lt(\Phi(\I),\Pi(\B)\Phi(\I)\rt)= \Psi_{\vp_{\eta}}(\B)
  \eeq
for any  $\B\in\AAA$.

In accordance with~\rr{12}, the functional $\Psi_{\vp_{\eta}}(\B)$
describes the mean of the observable $\B$ Â in the quantum state
$\Psi_{\vp_{\eta}}$. Equality~\rr{20} indicates that that this
mean is equal to the mathematical expectation of the operator
$\Pi(\B)$ in the state described by the vector  $\Phi(\I)$ in the
Hilbert space. This allows the full use of the mathematical
formalism of the standard quantum mechanics in calculating quantum
means in the proposed approach.

At the same time, the proposed approach differs essentially from
the standard quantum mechanics. In the latter, a relation of
type~\rr{20} is postulated (Born's postulate~\cc{born})and is the
starting point for constructing the so-called quantum probability
theory. But unlike the classical probability theory, the quantum
probability theory is not yet constructed as a nice mathematical
scheme. In this paper, formula~\rr{20} is derived as a consequence
of physically justified statements and of the classical
probability theory. In addition, we indicate when the formula is
valid: Eq.~\rr{20} is applicable for calculating means of
observables over a quantum ensemble.

\section {"Yes--no" experiment, collapse of a quantum state}

Let ${\cal R}=[-\infty,+\infty]$ be the extended real line and
${\cal S}$ (${\cal S}\subset{\cal R}$) be some interval or the
union of a denumerable number of intervals. Let some observable of
a physical system be measured in an experiment. We say that the
answer "yes" is obtained in the experiment if the measured value
of the observable is inside ${\cal S}$ and the answer "no" is
obtained if it is outside  ${\cal S}$. This procedure is called a
"yes--no" experiment. In fact, any real experiment reduces to
either a "yes--no" experiment or a sequence of such experiments.

It can be assumed that the value of a special observable quantity
is measured in a "yes--no" experiment. As such an observable, we
take a property of the system under investigation that produces
either the answer "yes" (in which case the observable is assigned
the value 1) or the answer "no" (in which case the observable is
assigned the value 0). If the measurement is reproducible, then
such an observable has the properties of a projector $\p$.   We
recall that a projector is any algebra element that satisfies the
conditions  $\p$ $\p^*=\p$ and $\p^2=\p$.

One of the central postulates in the standard quantum mechanics is
the so-called projection principle~\cc{von}. In its simplest form,
it can be stated as follows. Let a physical system be in a quantum
state described by a normalized vector $|\Psi\rangle$ of some
Hilbert space  \HHH. Let a "yes--no" experiment be performed (the
observable $\p$ measured) and the answer "yes" be obtained. As a
result of the experiment, the physical system then passes into the
new quantum state described by the vector
$|\Psi'\rangle=\p|\Psi\rangle[\langle\Psi|\p|\Psi\rangle]^{-1/2}$.

Equivalently, this statement can be expressed this way. The
vectors $|\Psi\rangle$ è $|\Psi'\rangle$ describe quantum states
corresponding to linear functional $\Psi(\A)=\langle\Psi|
\A|\Psi\rangle$ and $\Psi'(\A)=\langle\Psi'| \A|\Psi'\rangle$,
where $\A$ is any observable. We identify a quantum state with the
corresponding functional. As a result of the described experiment,
the physical system then passes from the quantum state $\Psi(\A)$
to the quantum state  $\Psi'(\A)$:

 \beq{21}
\Psi(\A) \to \Psi'(\A)=\Psi(\p\A\p)[\Psi(\p)]^{-1}.
 \eeq

The general form of the projection principle amounts to
postulating equality~\rr{21} for arbitrary quantum states, i.e.,
for states described by any linear positive normalized
functionals. The physical phenomenon governed by the projection
principle has been termed the collapse (reduction) of the quantum
state. The projection principle is broadly and successfully used
to describe the action of measuring devices on a quantum system.
At the same time, the collapse of a quantum state as a physical
phenomenon sharply contradicts our intuitive perceptions.

For example, a physical system that is in the state of an (almost)
plane wave is smeared over (almost) the entire coordinate space.
As a result of measuring a coordinate, this system is (almost)
instantly reduced to (almost) a point. There exist numerous
recipes for overcoming this paradox. But none of them seems
particularly convincing, to say the least.

The approach proposed here allows dropping the projection
principle as an independent postulate. Instead, it turns out to be
possible to obtain relation~\rr{21} as a corollary of the {\it
classical} probability theory applied to the statements formulated
above, which are not in conflict with physical intuition.

We assume that a physical system is in a quantum state $\Psi$
described by a linear functional $\Psi(\A)$. Let a device perform
a {\it reproducible} measurement of an observable $\p$, which is a
projector, and the measurement result be given by~1.

The considered physical system is an element of the corresponding
quantum ensemble. We imagine that a similar procedure is repeated
with other elements of this ensemble. As a result of such
measurements, the quantum ensemble corresponding to the quantum
state  $\Psi$ passes into another quantum ensemble. This new
ensemble corresponds to a quantum state $\Psi'$, and the means of
the observables  $\A$ over this ensemble are determined by the
functional  $\Psi'(\A)$.

This functional is obtained by the averaging procedure (see
formula~\rr{12}) of the functionals  $\vp$ corresponding to
elementary states. Each elementary state $\vp$ in the quantum
state $\Psi'(\A)$ is stable for the observable $\p$, and
$\vp(\p)=1$. In view of this and equality~\rr{12a}, the functional
$\Psi'$ satisfies the condition

 \beq{22}
  \Psi'(\p)=\Psi'(\I)=1.
  \eeq

Because $\Psi'(\A)$ is a positive linear functional, the
Cauchy-Bunyakovskii-Schwartz inequality holds for it. Therefore,

 \beq{23}
 |\Psi'(\A(\I-\p))|^2\leq\Psi'(\A^*\A)\Psi'(\I-\p).
\eeq
By virtue of  \rr{22}, the right-hand side of this inequality
is zero. Hence,

 \beq{24}
 \Psi'(\A)=\Psi'(\A\p).
 \eeq
Similarly,

 \beq{25}
 \Psi'(\A)=\Psi'(\p\A).
 \eeq
Replacing $\A\to(\I-\p)\A$ in \rr{24} and taking Eqs.\rr{24} and
\rr{25} into account, we obtain

 \beq{26}
\Psi'(\A)= \Psi'(\p\A\p)=\Psi'(\p\A\p)/\Psi'(\p).
 \eeq
Formula~\rr{26} implies that instead of finding the mean of $\A$
in the state  $\Psi'$, we can find the mean of the observable
$\A_p\equiv\p\A\p$.

Because  $[\A_p,\p]=0$, it follows that the observables  $\A_p$
and $\p$ are compatible. In view of formula~\rr{10}, the
probability distribution and hence the means of the observables
are independent of the type of measuring device used to determine
the values of these quantities. In our subsequent argument, we can
therefore restrict ourself to considering compatible measurements
of the observables $\A_p$ and $\p$. Such measurements must be
performed by devices belonging to a certain single type  $\eta$;
in the algebra of observables, this type then corresponds to a
maximal real commutative subalgebra $\QQ_{\eta}$ containing the
observables  $\A_p$ and $\p$.

We are interested in the means of the observable  $\A_p$ under the
condition that the value of  $\p$ is 1. We are not interested in
other observables, and instead of considering the full quantum
system, we can therefore restrict ourself to considering its
classical subsystem all of whose observables belong to the
subalgebra  $\QQ_{\eta}$. We can then use the rules for
calculating a conditional mean in classical probability theory.

We recall that the conditional probability  $P(F({\cal S})|F(p))$
of an event  $F({\cal S})$ under the condition that the event
$F(p)$ has occurred is given by (see, e.g.,~\cc{nev})

  \beq{27}
P(F({\cal S})|F(p))=\frac{P(F({\cal S})\cap F(p))}{P(F(p))},
 \eeq
where  $P(F(p))$ is the probability of $F(p)$ (it is assumed that
$P(F(p))\neq 0$), and $P(F({\cal S})\cap F(p))$ is the probability
of a simultaneous occurrence of the events  $F({\cal S})$ and
$F(p)$.

In the case under consideration, the role of the space of
elementary events is played by the set of elementary states
belonging to the quantum state  $\Psi$. The event  $F(p)$
corresponds to the set of those elementary states (elementary
events) for which  $\vp_{\eta}(\p)=1$. The event  $F({\cal S})$
corresponds to the set of elementary states for which
$\vp_{\eta}(\A_p) \in {\cal S}$, where ${\cal S}$ is the subset of
the real axis described at the beginning of this section, and
$F({\cal S})\cap F(p)$ is the intersection of the sets  $F({\cal
S})$ and $F(p)$.

Because
$\vp_{\eta}(\A_p)=\vp_{\eta}(\p\A_p)=\vp_{\eta}(\p)\vp_{\eta}(\A_p)$,
it follows that  $\vp_{\eta}(\A_p)=0$ if $\vp_{\eta}(\p)=0$.
Therefore, the mean of  $\A_p$ evaluated using the probabilistic
measure  $P(F({\cal S})\cap F(p))$ coincides with the mean
evaluated using the measure $P(F({\cal S}))$. On the other hand,
because  $\vp_{\eta}(\p)$ is equal to either  1 or 0, it follows
that  $P(F(p))$ is equal to the mean of the observable  $\p$.

Therefore,~\rr{27} implies that the conditional mean of $\A_p$ is
given by
 $$
 \langle\A_p|F(p)\rangle=\frac{\Psi(\A_p)}{\Psi(\p)}=
 \frac{\Psi(\p\A\p)}{\Psi(\p)}.
 $$

By its physical meaning, the quantum mean  $\Psi'(\p\A\p)$
involved in~\rr{26} coincides with the conditional mean
$\langle\A_p|F(p)\rangle$. Therefore, formula~\rr{26} can be
rewritten as
 \beq{29}
 \Psi'(\A)= \frac{\Psi(\p\A\p)}{\Psi(\p)},
 \eeq
and we thus obtain formula~\rr{21}, which is the mathematical
formulation of the projection principle in the standard quantum
mechanics. Unlike formula~\rr{21} formula~\rr{29} is a corollary
of only intuitively comprehensible statements and of the classical
probability theory. We also note that formula~\rr{29} pertains to
a quantum ensemble (quantum state), not to an individual term of
this ensemble (elementary state). Therefore, when we say, for
example, that some quantum particle is in a plane-wave state, this
means that its elementary state belongs to the set of elementary
states for which the momentum value is the same but the coordinate
values are different for different terms in the ensemble. This
does not mean that the particle is smeared over the entire
coordinate space. An elementary state of a concrete particle can
be such that the coordinate values are localized in a sufficiently
small domain. There can be many of these values because this
elementary state is not necessarily stable with respect to the
coordinate. In measuring the coordinate of this particle, we do
not contract the plane wave to a point. We simply determine a
localized coordinate domain for the elementary state of the
particle. Obviously, as a result of the action of the measuring
device, the elementary state of the particle changes. First, if
the measurement is reproducible, the state becomes stable with
respect to the coordinate. Second, the momentum value changes. But
this does not mean that the particle becomes smeared in the
momentum space. Just its elementary state becomes a term of
another quantum ensemble, which contains elementary states
corresponding to different momentum values.

\section{Conclusions}

\begin{figure}
\begin{center}
\includegraphics{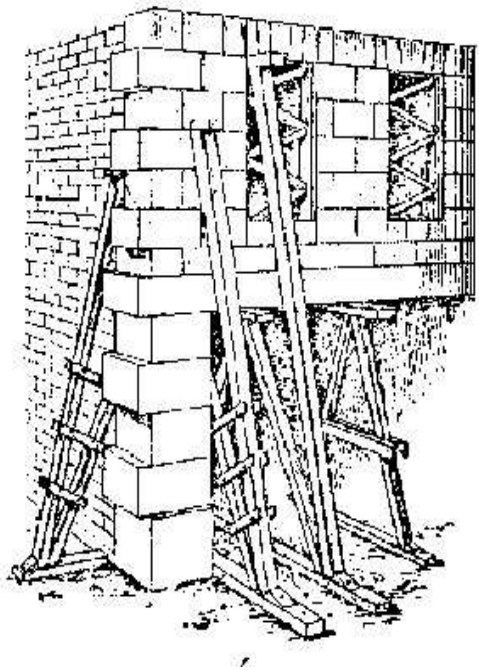}
\caption{\label{fig1} Shoring the edifice (reproduced from Fig.
813 in \cc{bar}).}
\end{center}
\end{figure}

The approach to quantum theory described in this paper by no means
disproves the standard quantum mechanics. The founding fathers of
quantum mechanics erected a remarkable edifice. But they began the
construction with the second floor, the description of
probabilities and means. Therefore, the stability of this edifice
has required a large amount of shoring (see Fig. 1) in the form of
a series of "principles": the superposition principle, the
uncertainty principle, the  principle of complementarity, the
projection principle, the indistinguishability principle, and the
principle of the absence of trajectories. All these principles
appear rather artificial and are not easily amenable to physical
interpretation. The main task of these principles is to justify
the mathematical formalism of the standard quantum mechanics.
True, this mathematical formalism has proved amazingly
serendipitous, but this is not the case with its physical
interpretation. It is not without reason that discussions of the
physical interpretation of quantum mechanics still vividly
proceed, although the term "physical interpretation" itself seems
quite strange. If quantum mechanics is a physical theory, then it
must not need any physical interpretation. By using this term, we
admit, be it willingly or not, that quantum mechanics is not a
physical theory but a mathematical model. In this work, we have
attempted to construct quantum mechanics just as a physical
theory, based on experimental data.

The central point of the described approach is the introduction of
the notion of an "elementary state," which is absent in the
formalism of the standard quantum mechanics. This notion, on one
hand, gives a clear mathematical counterpart of such a physical
phenomenon as an individual experimental act. On the other hand,
it allows using the well-developed formalism of classical logic
and classical probability theory. It must be borne in mind here
that although the references to the so-called quantum logic and
quantum probability theory may be rather frequent, it has so far
been impossible to give them the structure of a clear-cut complete
theoretical scheme.

Based on the notion of an elementary state and using the {\it
classical} probability theory, we can completely reproduce the
mathematical formalism of the standard quantum mechanics and
simultaneously show its applicability domain. This formalism
applies to quantum ensembles. This is a very important type of
ensemble but not the most general one by far. In particular, this
formalism is not suitable for describing an individual event.

\end{document}